# UTA versus line emission for EUVL: Studies on xenon emission at the NIST EBIT

K. Fahy, P. Dunne, L. McKinney, G. O'Sullivan, E. Sokell and J. White
Department of Experimental Physics, University College Dublin, Belfield, Dublin 4, Ireland.
A. Aguilar, J. M. Pomeroy, J. N. Tan, B. Blagojevi_, E.-O. LeBigot[1] and J. D. Gillaspy.
Atomic Physics Division, National Institute of Standards and Technology, Gaithersburg, MD 20899, USA
[1] Université P and M Curie et Ecole Normale Supéieure, 75252 Paris, France.

**Abstract:**
Spectra from xenon ions have been recorded at the NIST EBIT and the emission into a 2% bandwidth at 13.5 nm arising from 4d_5p transitions compared with that from 4d_4f and 4p_4d transitions in Xe XI and also with that obtained from the unresolved transition array (UTA) observed to peak just below 11 nm. It was found that an improvement of a factor of five could be gained in photon yield using the UTA rather than the 4d_5p emission. The results are compared with atomic structure calculations and imply that a significant gain in efficiency should be obtained using tin, in which the emission at 13.5 nm comes from a similar UTA, rather than xenon as an EUVL source material.

## 1 BACKGROUND:

The search for the optimum radiation source at 13.5 nm is one of the main challenges in EUV physics today. Currently the quest is to find a source with a conversion efficiency (CE) of 3% into 2% bandwidth as defined by the needs of the microelectronic industry [1]. The earliest work was performed with laser produced plasmas generated on solid targets. Kauffman *et al.* [2] attained a CE close to 1% into 3% bandwidth using 7.5 ns, 300 mJ frequency doubled Nd:YAG pulses focussed to a power density of $2\times10^{11}$ Wcm$^{-2}$ onto a tin target. Subsequently Spitzer *et al.* [3] obtained a CE > 0.8% into 3% bandwidth, again using tin irradiated at power densities of $1$-$2\times10^{11}$ Wcm$^{-2}$. In their work they performed an extensive survey of the emission from a large range of elements and found that the emission at the required wavelength peaked near tin. At higher power densities they noted that the intensity decreased. Shevelko *et al.* [4] also undertook an extensive study of the spectra of a large number of elements from laser produced plasmas with a KrF excimer laser focussed to a power density of $10^{12}$ Wcm$^{-2}$ onto planar targets including tin and found that the maximum intensity was in fact obtained for Ge and Re under these conditions. So early work already pointed to the sensitivity of the emission of specific elements to laser power density.

In order to avoid the debris problems associated with solid targets other target materials were sought with strong emission in the 13-14 nm region. Jin and Richardson [5] used mass limited water ice targets, where the emission is from the O VI lines near 13 nm, to limit the debris. However, all of their early work with mass-limited targets produced significant levels of debris both in the form of ions and particulates of various sizes. The need to reduce particulate emission led to the choice of xenon. There is a line group between 13-14 nm in the spectrum of Xe which has been shown by a number of researchers to arise from $4d^8$ _ $4d^75p$ transitions in Xe XI [6]. These lines have recently been identified by Churilov *et al.* [7] by comparison of new very high resolution data with atomic structure calculations using the suite of codes developed by Cowan [8]. Considerable work has been expended on exploring the feasibility of using laser produced plasmas of xenon clusters produced by supersonic jets or gas puffs from nozzles or solid xenon targets [9]. The highest conversion efficiencies (1.2% into 2% bandwidth) have been achieved using solid xenon by Shields *et al.* [10]. Liquid jet targets were also developed and conversion efficiencies of the order of 0.75% have been reported with these [11]. In addition, a wide variety of pulsed discharge sources using xenon or xenon/helium mixtures are being actively investigated. A conversion efficiency of 0.8% into a $2\pi$ solid angle has recently been reported [12].

As already stated the relevant emission from xenon originates from a single ion stage, $Xe^{10+}$, for which a theoretical spectrum, calculated with the Cowan suite of codes is shown in Fig. 1. From this diagram it is obvious that the $4p^64d^8$ _ $(4p^54d^9 + 4d^74f)$ line group near 11 nm is significantly stronger than the 4d _ 5p transitions occurring near 13.5 nm. In addition, plasma modelling predicts that the maximum concentration of $Xe^{10+}$ attainable is only of the order of 50%. So any advantages of using xenon, based on its gaseous nature and inherent cleanliness, must be weighed up against the disadvantages of a low density of emitting ions and the fact that the transitions involved are at least a factor of ten weaker than those occurring in the 11 nm band. Moreover the 13.5 nm wavelength of choice almost coincides with that of maximum absorption in the neutral and low ion stages [13].

If instead of xenon, one considers tin, the emission arising from $4p^64d^n$_$(4p^54d^{n+1} + 4d^{n-1}4f)$ transitions in a range of adjacent ion stages overlap to produce an unresolved transition array (UTA) centred near 13.5 nm [14]. Final state configuration interaction between the $4p^54d^{n+1}$ and $4d^{n-1}4f$ configurations causes a narrowing of the array and also causes the most intense lines to essentially overlap in energy [15-17]. The profile of the UTA in this type of transition is known to be sensitive to concentration of the element of interest because of opacity effects [18]. O'Sullivan and Faulkner [14] showed that if the tin concentration was reduced to approximately 10%, the peak brightness actually increased. This result also agrees with the data of Choi *et al.* [19] who, in a space resolved study of emission from tin metal and tin oxide targets, found that the in band emission from a $SnO_2$ film target was greater than that from a tin slab target. For mass limited targets, the peak intensity essentially is unaltered but the profile width decreases as the tin concentration is reduced [20]. Thus optical opacity effects are extremely important and radiation transport effects must be included in any attempt to model the plasmas. Furthermore, if the remaining 90% of the target constituents are low Z materials the radiation emitted is concentrated in a band 1-2-nm wide centred near 13.7 nm and the ordinary recombination continuum from the plasma is suppressed in comparison with that from a pure tin target. In fact if pure tin targets are used the recombination continuum contains most of the emitted radiation. It extends throughout the XUV and if absorbed by the multilayer optics could cause heating, possible distortion and loss of focus. These results indicate that a composite target, including approximately 10% tin, would be an appropriate candidate for a 13.5 nm source.

The large in band intensity emitted from tin plasmas has encouraged their study and CE values of 2% or greater have been reported [11,21,22]. The maximum CE attainable is expected to exceed the 3% target set by industry. However theoretical determination of the maximum CE attainable is hampered by a lack of experimental data. For example no identification exists for any of the strong lines within the UTA. Indeed, the wavelength ranges at which different ions emit most strongly within the overall UTA envelope are not known either.

Under certain conditions, UTA involving $\delta n = 0$ transitions are the strongest features observed in EUV spectra from laser produced plasmas [23]. To understand why, one needs to look at the redistribution of oscillator strength from the 4d _ εf resonance in the neutral spectrum to 4d _ 4f transitions in the ions. In neutral tin the effective interaction between the nucleus and an $l=3$ electron contains attractive coulomb and repulsive centrifugal terms that result in a bimodal potential with two pronounced minima. The first minimum, with is centre near the Bohr radius, $a_0$, is close to the 4d wavefunction peak but lacks sufficient depth to support a bound level and, as a result, the 4d → f photoexcitation channel is dominated by a 4d → εf shape resonance that forms when the outgoing f electron has sufficient energy to overcome the centrifugal barrier. The outer shallow well, centred near 16 $a_0$, is essentially hydrogenic in character and contains the 4f level. Because there is essentially no overlap with the 4d wavefunction, 4d→4f lines are absent [24].

However, with increasing ionisation the centrifugal barrier disappears and 4f wavefunction contraction into the inner well region results in a transfer of oscillator strength from the continuum to the discrete spectrum. The inner potential well becomes deeper and narrower and the *4f* wavefunction gradually contracts into the core region where it has a large spatial overlap with the *4d* wavefunction. Consequently, 4d _ 4f transitions become intense and essentially contain all of the oscillator strength associated with the neutral shape resonance. These neutral resonances, where they occur, completely dominate the EUV photoabsorption spectra so the transitions that evolve from them, the $\delta n = 0$ UTA, likewise dominate EUV emission spectra. This evolution has been studied experimentally for a number of isonuclear sequences [25,26] and in each case the transfer of oscillator strength from continuum to discrete features is clearly seen.

This paper seeks to address the latter point by using Xe spectra acquired at the NIST EBIT, which produces an optically thin plasma and allows some degree of ion stage differentiation, to gain information on the contribution of the emission of individual ion stages to the UTA at 11 nm and enable comparison of the emission intensity to that at 13.5 nm. These can then be used to calibrate atomic structure calculations which, when extended to tin, should provide useful information on the spectral contribution of different ion stages to the UTA and shed some light on the maximum CE attainable.

**Experiment:**

Spectra were obtained at the NIST EBIT (Electron Beam Ion Trap) [27] using a xenon injection pressure between $5 \times 10^{-5}$ torr and $9 \times 10^{-5}$ torr and a range of accelerator and trap voltages. EBITs were developed to perform spectroscopic studies of highly charged ions, which are created by sequential electron impact ionisation. Once positively charged ions are formed they are confined to the trap region by a mixture of magnetic and electrostatic fields. Radiation emitted from the plasma within the trap is viewed through side windows at 90 degrees to the electron beam. The experiments involved varying EBIT operating parameters, such as the beam energy (which determines the highest charge state that can be created), so that the distribution of ion stages within the plasma changed systematically. The spectra were recorded on a 0.25 m flat field grating vacuum spectrograph equipped with a variable line space grating of the sort described by Kita *et al.* [28] and fitted with a CCD [29]. Individual spectra were integrated typically over five minute intervals. The spectral resolution varied from 0.02 at 5 nm to 0.035 at 19 nm and the spectra were corrected for cosmic ray events.

For comparison, spectra of laser produced plasmas formed on solid xenon slab targets were recorded with a 2.2 m Schwob-Fraenkel spectrograph with mean resolution of 0.001 in the 11 to 14 nm region. A Nd:YAG laser operating at $\lambda=1.06$ µm and capable of delivering up to 1 J in a 20 ns pulse was used to generate the EUV emitting plasmas. The output was focused typically to a spot size of diameter ~ 100 µm. The maximum average power density attained was $1 \times 10^{12}$ W cm$^{-2}$. According to the collisional radiative equilibrium (CR) model of Colombant and Tonon [30] the corresponding electron temperature was close to 200 eV.

**Theoretical:**
Using the Cowan suite of codes, it is possible to calculate theoretical spectra (gA values versus $\lambda$) for a range of ion stages in xenon. The results are shown in Fig. 2 for ion stages IX (ground configuration $4d^{10}$) through XVIII (ground configuration $4d^{1}$), i.e. those stages containing 4d electrons in their valence or outermost subshell, for xenon. In each of these spectra the bulk of the resonance emission arises from $4p^{6}4d^{n}$ _ $(4p^{5}4d^{n+1} + 4d^{n-1}4f)$ lines which merge to form a UTA lying in the 10–12 nm region in xenon. This UTA occurs in the 13-14 nm region in tin. Final state configuration interaction between the $4p^{5}4d^{n+1}$ and $4d^{n-1}4f$ configurations causes these transitions to overlap in energy [15].

In an optically thin plasma, these lines should completely dominate the EUV emission. At the ion densities encountered in laser produced plasmas and also in discharge plasmas, opacity effects become important and have the effect of reducing the intensity of the UTA by comparison with that of the 4d _ 5p transition groups lying at longer wavelengths. For example, one such group in Xe X at 15 nm, identified by Kaufman and Sugar [31] and Churilov and Joshi [32], was recently found by Böwering *et al.* [33] to be particularly intense under a particular set of experimental conditions in a discharge plasma. No allowance for opacity effects has been made in the spectra shown in Figs. 1 and 2. Recently calculations have been preformed using the HULLAC code [34] to allow for radiation transport effects in xenon [35]. The results show the correct intensity trends but because the line positions are derived from ab initio calculations the peaks are greatly displaced from their true positions.

Moreover, because of the high densities encountered in laser produced plasmas emission from satellite lines must also be included in any complete model. The impact of such lines on the xenon emission spectrum has recently been reported by Sasaki [36].

In contrast in the case of EBIT spectra, opacity effects are negligible and the spectra might be expected to mirror the calculated oscillator strength differences better than emission spectra from laser produced plasmas. At the electron densities appropriate to the EBIT spectra presented here, ($10^{13}$-$10^{14}$ cm$^{-3}$), collisional excitation rates must be taken into account [37] These considerations lead to considerable differences in the calculated spectra especially for very highly charged ions. This is clearly evident from the spectra shown in Fig. 3 which can be compared with that from a laser produced plasma (Fig. 4), where absorption features in Xe VI, VII, IX, [38] are clearly evident.

**Results**

From Fig. 1, it is seen that the summed gA value available in 4d _ 5p transitions is significantly less than that attainable in the UTA, corresponding to $4p^6 4d^8$ _ ($4p^5 4d^9$ + $4d^7 4f$). In order to compare the relative intensities of the δn = 0 transitions at 11 nm with those at 13.5 nm in Xe XI, it is useful to compare the summed gA values for transitions falling within a 2% bandwidth at 13.5 nm and a similar width window (0.27 nm) centred at 11.17 nm, which corresponds to the emission peak of the Xe XI UTA [39]. In Fig. 5 an EBIT emission spectrum obtained at a beam energy of 210 eV (the ionisation potential required to get 11 times ionised xenon is 226 eV) is compared with the high resolution experimental data of Churilov *et al.* [39], which has been convolved with an instrument profile with a full width half maximum of 0.02 nm (assumed to be gaussian for simplicity) pertinent to the detection system employed on the EBIT. Once the two spectra have been scaled to the same peak intensity at 11.17 nm, the agreement between the two spectra is good, suggesting that Xe XI, the most highly charged ion stage obtainable, is the strongest emitter in this wavelength region present in the EBIT under these experimental conditions. The number of counts within the bandwidths of interest were 37.3/s at 11.17 nm and 11.9/s at 13.5 nm, giving a ratio of 3 (see Table 1). In taking this ratio a constant background obtained from the tails of the spectrum was assumed at each location. This experimental ratio can be compared with a value of 1.4 obtained from the vacuum spark data of Churilov *et al.* [39], assuming a resolving power of fwhm = 0.02nm. In the case of the EBIT data, the system response was essentially flat across the wavelength range of interest, while the response of the system used by Churilov *et al.* [39] has not been quantified. If one assumes that the response of this system does not vary greatly from 11 to 13.5 nm, then it seems likely that opacity effects could be responsible for the different values. The corresponding ratio obtained from calculations was 20. The ratio is particularly sensitive to the scaling factors applied to the Slater-Condon parameters in the calculations. Using scaling parameters of 90% for $F^k$ and $G^k$ the ratio was 4 while using 80% (4d_4f) and 85% (4d_5p), the values that best fit the data of Churilov *et al.* [39], gave the value quoted above. The difference arises from the fact that a change of scaling of only 1% causes a shift in wavelength of 0.1 nm, which is sufficient to move the strongest lines in or out of the window chosen for the summation. This is further borne out by the fact that if both calculated summations are performed over a 7.5% bandwidth, centred at 13.5 nm and 11.17 nm respectively, the ratio is 15 in both cases. The difference between the theoretical and experimental values is most likely to be due to the assumption that the line emission is determined by the gA values and no consideration has been made for variations among the population of the emitting levels or collisional effects. The sensitivity of the calculations to the scaling factors might also contribute to the discrepancy between the measurements and calculations. Nonetheless the experimental results obtained here confirm that a source based on emission associated with δn = 0 transitions will lead to higher CE values.

A spectrum recorded at an EBIT beam energy of 1000 eV is shown in Fig. 6. The enhancement of the shorter wavelength features in comparison to the sharp feature at 12 nm is evident from the figure. This sharp feature results from emission from Xe IX [32]. The lower wavelength side of the δn = 0 UTA is also enhanced in the 1000 eV spectrum in comparison with that recorded at a beam energy of 500 eV (Fig . 3). There are also features in the 13.5 nm band that are not evident in the 500 eV spectrum. These observations are consistent with the production of higher charge states as the energy of the electron beam is increased.

In the EBIT spectra, the observed UTA peaks at an energy of 10.84 nm, which is slightly higher than the observed emission peak of Xe XI at 11.17 nm [39]. The count rates within a 0.27 nm band centred here are compared with the counts at 13.5 and 11.17 nm in table 1. In Fig. 7 we plot the two in band ratios as a function of EBIT beam energy. This plot clearly shows the effect of additional ion stages at lower wavelength. The maximum ratio of the emission from the 10.84 nm band to the 13.5 nm band is 5.2, obtained at an EBIT energy of 500 eV. The large increase with beam energy for the band centred at 10.84 nm is consistent with the increase being associated with emission from higher charge states at lower wavelengths. The calculations shown in Fig. 2 indicate that charge states from $Xe^{10+}$ to $Xe^{17+}$ exhibit δn = 0 UTAs between 10 and 11 nm, although there are relatively few lines in the $Xe^{17+}$ feature. The relatively constant ratio associated with the 11.17 nm band is due to the fact that the emission in both of the relevant areas of the spectrum is dominated by Xe XI transitions. At higher EBIT beam energies both ratios decrease due to contributions to the 13.5 nm region of the spectrum from ion stages that start to appear in the plasma.

The comparison between our model and the EBIT experimental data suggests that similar calculations coupled with measurements in tin could be used to estimate the amount of emission that will occur at 13.5 nm from the analogous UTA. Ion stages from $Sn^{8+}$ through $Sn^{12+}$ will contribute to the δn = 0 UTA. If one considers the spectrum of Sn VII, which is isoelectronic with Xe XI, and takes the ratio of 4d _ 5p to 4$l$ _ 4$l$ emission again within a 0.27 nm window centred on the peak emission of each, a ratio of 9 is obtained. This compares with a ratio of 20 in xenon, this reduction is to be expected as the 4d 4f overlap in Sn VII is significantly less than that in Xe XI due to the charge dependent nature of the 4f contraction [14,38].
If one assumes that the same mechanisms responsible for the disparity between theory and experiment in xenon are operative in the case of tin, then the observed ratio, which is 5.2 in xenon, would be reduced to 2.3 in tin. Combining the measured ratio of 5.2 with a recently recorded CE at 13.5 nm in xenon [10], suggests that the maximum in band CE that may be derived from a xenon plasma source could be close to 6% at a wavelength of 10.84 nm. However, plasma opacity effects may influence this figure.

**Conclusion:**

We have shown that the emission intensity available from the UTA appearing in the spectrum of ionised xenon at 11 nm in a low electron density regime is significantly greater than that obtained at 13.5 nm, the wavelength of choice for EUVL. In a comparison between atomic structure calculations and experiment we have found that the calculations overestimate the ratio by a factor close to seven. Even allowing for this disparity, we estimate that the intensity enhancement obtained from moving to tin as the source material of choice would lead to an enhancement of more than a factor of two compared to xenon. We intend to investigate the emission from tin in the near future in order to identify the contributions of different ion stages and their range of influence on the UTA occurring in tin plasmas near 13.5 nm.

**ACKNOWLEDGEMENT:**
This work was supported by Science Foundation Ireland under Investigator Grant **02/IN.1/I99**, International SEMATECH under LITH152, and the U.S. Department of Commerce

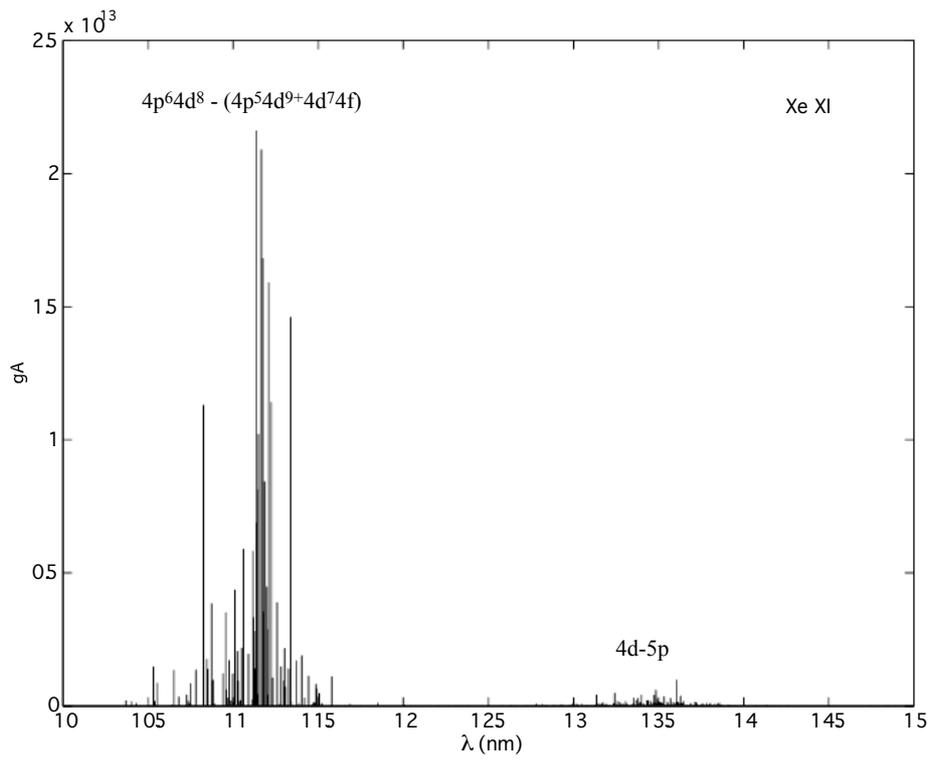

Fig 1: Theoretical oscillator strength (gA-value) spectrum for $Xe^{10+}$.

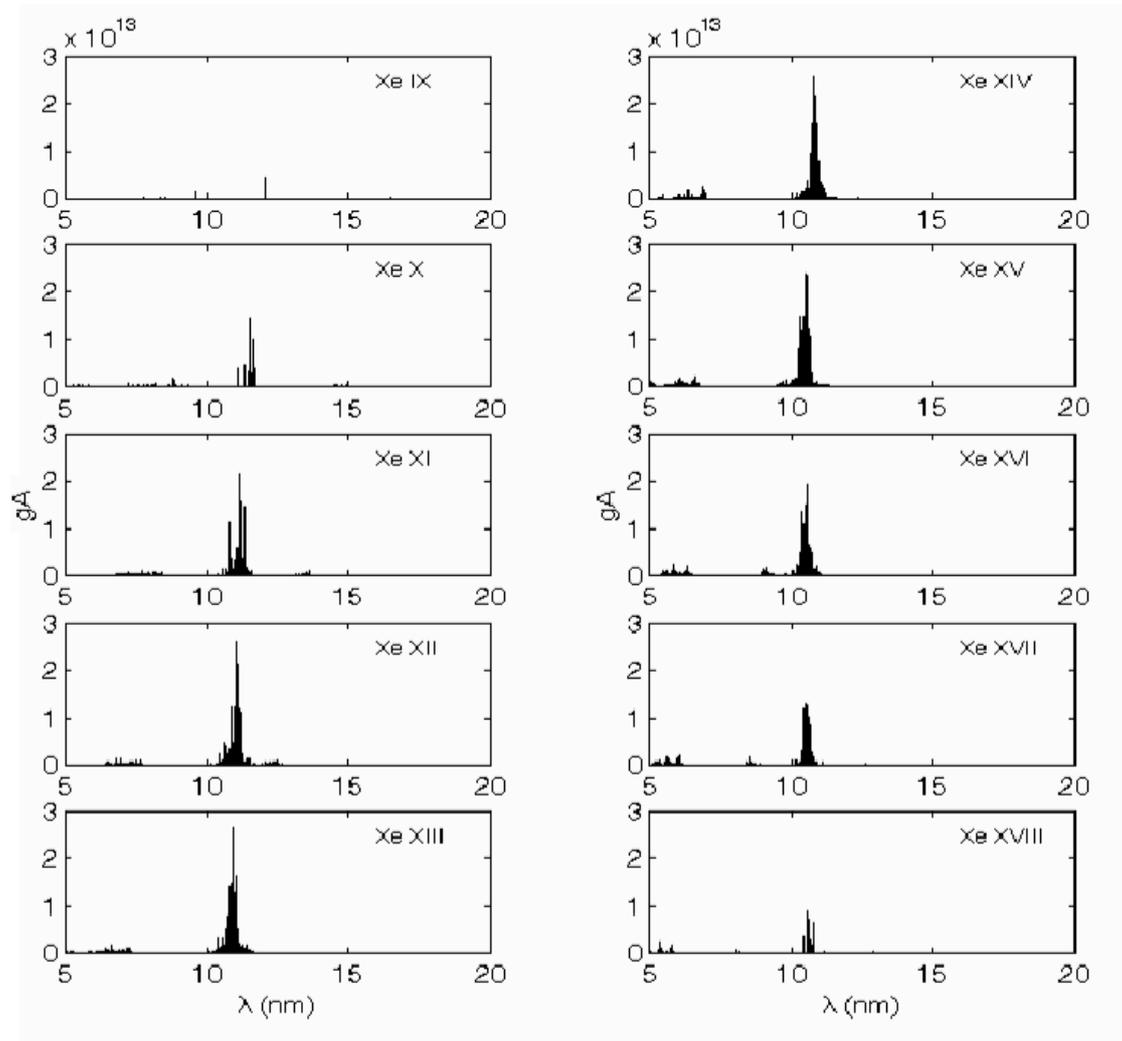

Fig. 2: Theoretical plots of oscillator strength (gA-value) versus wavelength for xenon. All ion stages with valence 4d electrons are shown. The plots illustrate the relative weakness of the longer wavelength 4d_5p transitions, which are responsible for the emission at 13.5 nm in Xe XI.

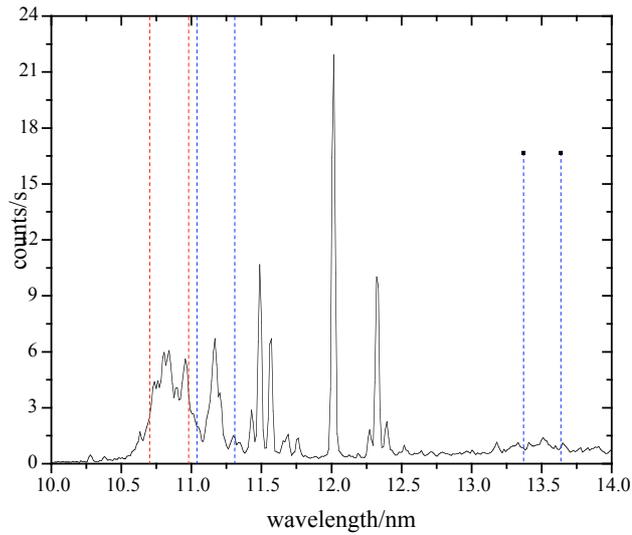

Fig. 3: EBIT spectrum of xenon obtained at a beam energy of 500 eV and a beam current of 1.2mA. The spectrum was accumulated over 60 minutes and the xenon injection pressure was $5 \times 10^{-5}$ torr. Vertical lines indicate the 0.27nm bandwidths referred to in the text.

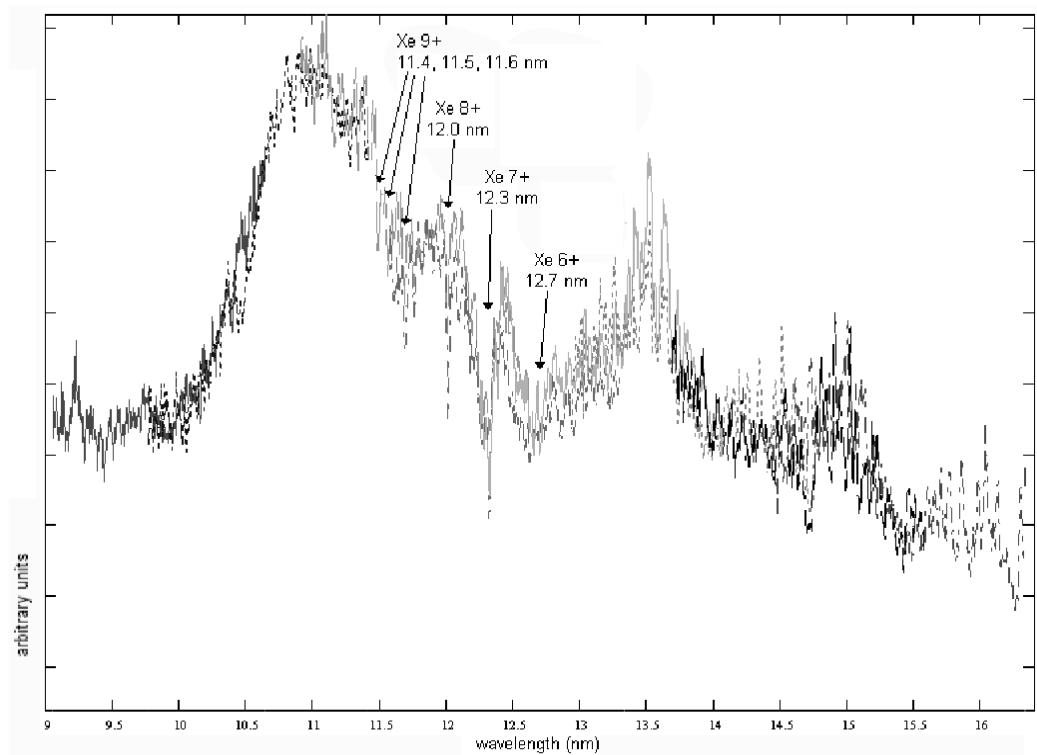

Fig. 4: Laser produced plasma spectrum from a solid xenon target. Note that the UTA is significantly broader than in the EBIT spectrum and there is strong underlying recombination continuum emission. Opacity effects are clearly evident in the spectrum and major absorption features arising from lower ion stages are indicated on the plot. Note also that because of opacity the peak intensities at 11 nm and 13.5 nm are closer than in the optically thin EBIT spectrum. The spectrum is comprised of data obtained at different detector positions, indicated by the different line styles.

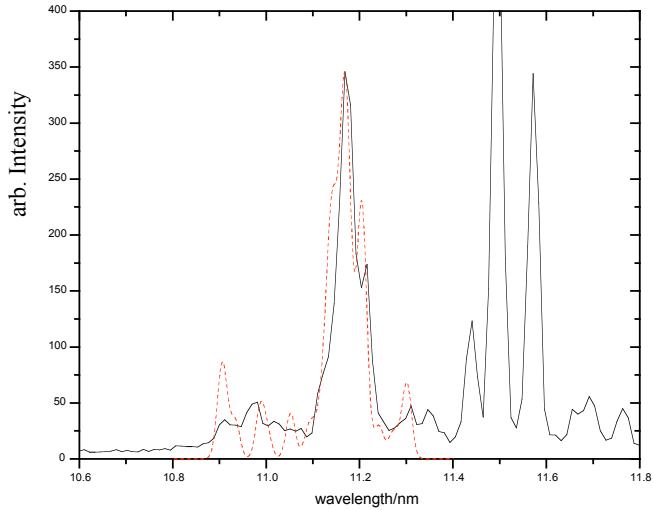

Fig. 5. EBIT spectrum of xenon obtained at a beam energy of 210 eV and a beam current of 0.4 mA. The spectrum was accumulated over 120 minutes and the xenon injection pressure was $9\times10^{-5}$ torr. Under these conditions Xe XI is the dominant ion stage in the EBIT plasma. The dashed line is a synthetic spectrum that has been created by convolving the high resolution Xe XI data of Churilov et al 2004 with an appropriate instrument function. The synthetic spectrum has been scaled to match the experimental spectrum intensity at 11.17 nm.

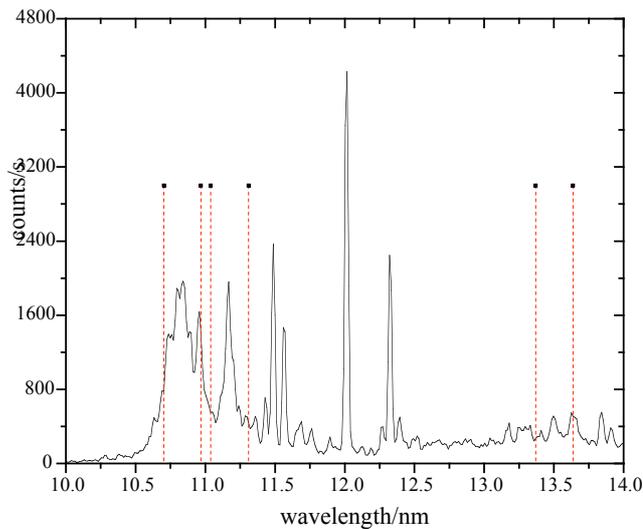

Fig. 6: EBIT spectrum of xenon obtained at a beam energy of 1000 eV and a beam current of 6mA. The spectrum was accumulated over 60 minutes and the xenon injection pressure was $5\times10^{-5}$ torr. Vertical lines indicate the 0.27nm bandwidths referred to in the text

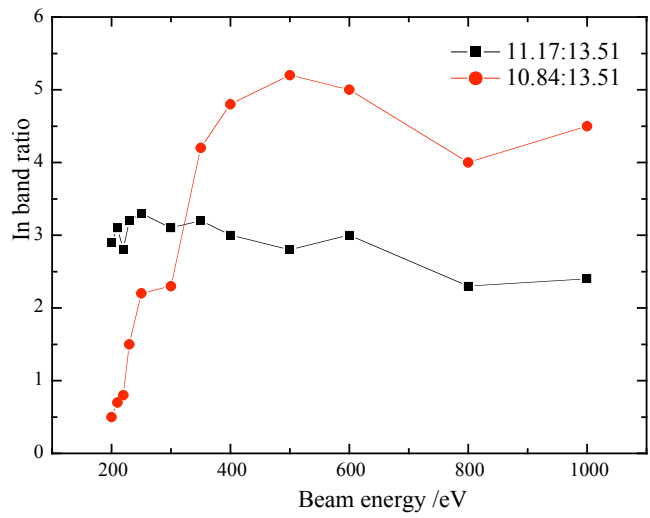

Fig. 7: In band ratios as a function of EBIT beam energy. The ratios are obtained by dividing the number of counts in a 0.27 nm band centred at 11.17 (squares) and 10.84 (circles) by the counts in a similar band centred at 13.51nm.

Table 1: In band counts and ratios as a function of EBIT energy. The values for the total counts were obtained by summing all of the counts that occurred within a 0.27nm bandwidth centred at the indicated energy, after a constant background had been removed from each spectrum.

| E/eV | Total counts/s | | | Ratio's | |
|---|---|---|---|---|---|
| | 13.51nm | 11.17nm | 10.84nm | 11.17:13.51 | 10.84:13.51 |
| 200 | 11.4 | 33.3 | 5.3 | 2.9 | 0.5 |
| 210 | 11.9 | 37.3 | 7.8 | 3.1 | 0.7 |
| 220 | 6.0 | 16.9 | 4.8 | 2.8 | 0.8 |
| 230 | 7.9 | 24.9 | 11.6 | 3.2 | 1.5 |
| 250 | 7.2 | 23.5 | 15.9 | 3.3 | 2.2 |
| 300 | 3.8 | 11.8 | 8.9 | 3.1 | 2.3 |
| 350 | 7.1 | 22.6 | 29.6 | 3.2 | 4.2 |
| 400 | 10.9 | 32.3 | 51.8 | 3 | 4.8 |
| 500 | 21.6 | 60.9 | 111.7 | 2.8 | 5.2 |
| 600 | 26.4 | 78.3 | 131.1 | 3 | 5 |
| 800 | 88.8 | 202.4 | 353.0 | 2.3 | 4 |
| 1000 | 135.0 | 319.4 | 604.9 | 2.4 | 4.5 |